\documentclass{elsart}

\usepackage{graphicx} 

\usepackage{float} 

\usepackage{citesort}

\usepackage{amssymb}

\begin{document}

\begin{frontmatter}















\title{Appropriateness of correlated first order auto-regressive processes for modeling daily temperature records}





\author[adr1]{Radhakrishnan Nagarajan}
\author[adr2]{R.B. Govindan\corauthref{cor1}},
\corauth[cor1]{R.B. Govindan,
Graduate Institute of Technology,
University of Arkansas at Little Rock,
Little Rock, AR 72204}
\ead{rbgovindan@git.ualr.edu}
\address[adr1] {Center on Aging, University of Arkansas for Medical Sciences, Little Rock, AR 72205, USA}

\address[adr2] {Graduate Institute of Technology, University of Arkansas, Little Rock, AR 72204, USA}

\begin{abstract}
The present study investigates linear and volatile (nonlinear) correlations of first-order autoregressive 
process with uncorrelated AR (1) and long-range correlated CAR (1) Gaussian innovations as a function 
of the process parameter ($\theta$).  In the light of recent findings \cite{jano}, we discuss the choice of
CAR (1) in modeling daily temperature records. We demonstrate that while CAR (1) is able to capture 
linear correlations it is unable to capture nonlinear (volatile) correlations in daily temperature records.
\end{abstract} 


\end{frontmatter}


\section{Introduction} 


Noise at the dynamical and measurement level often encourages the choice of stochastic
 interpretation of time series data. Detrended fluctuation analysis (DFA) \cite{peng1,peng2} and its extensions \cite{ya1,jk1,ay2}
have been used widely to determine the nature of correlations from stochastic processes obtained from 
a wide-range of systems \cite{gas}. These studies have subsequently suggested the choice of stochastic 
models to explain complex behavior observed in synthetic and natural data sets \cite{jano}. The objective 
of the present study is to understand the scaling behavior of the first-order stochastic processes and 
their volatility series \cite{ya1} as function of the process parameter. Based on the results obtained for the 
various parameter regime, we evaluate the choice of these processes in modeling real-world temperature 
data \cite{jano}.
\section{First order auto-regressive process and its extension}
\subsection{AR (1) with uncorrelated Gaussian innovations}
Classical AR (1) is given by the expression
$$ x_n=\theta x_{n-1}+\epsilon_n, n=1..N,  \eqno(1) $$
where $\epsilon_n$ represents independent and identically distributed (i.i.d)
Gaussian innovations with zero mean and unit variance.
 The corresponding volatility series of (1) is given by $y_n=\vert x_n-x_{n-1}\vert$, where $\vert w \vert$  represents the absolute value of $w$. 
Correlations in $y_n$ imply that clusters of (small) big changes are likely to be
followed by clusters of (small) big changes. It can be shown analytically that
Eqn. (1) is stationary when $\vert \theta \vert <1$ i.e. $-1<\theta<1$. The autocorrelation function 
(ACF) of (1) decays exponentially and is of the form $\rho(k)=\theta^k$ for lag $k$. The nature of the decay is governed by the 
process parameter $\theta$. While $\rho(k) = \theta^k$ exhibits an exponential decay for ($0<\theta<1$), it exhibits an exponential decay with 
oscillations for ($-1<\theta<0$). It is important to note that $(-1<\theta<0)$  and $(0<\theta<1)$ represent cases where the current sample is 
negatively and positively correlated with its immediate past, respectively. The strength of the short-term correlated and short-term 
anti-correlated behavior is dictated by the process parameter $\theta$.

\subsection{AR (1) with long-range correlated Gaussian innovations, CAR (1)}

We also consider the case where $\epsilon_n$ in (1) is generated by a
long-range correlated Gaussian process \cite{makse}. This extended AR (1) with 
long-range correlated Gaussian innovations shall be referred to as CAR (1) in the subsequent sections. The choice of the term CAR 
was encouraged by recent studies \cite{jano}.  In the light of recent findings
\cite{jano}, we critically evaluate the choice of CAR (1) for modeling
real-world daily temperature records.

\section{Results and discussion}
A complete description of DFA and its extensions can be found elsewhere \cite{peng2,ya1,jk1}. In the present study, the length of the data sets is 
chosen sufficiently larger $N =2^{14}$, in order to avoid bias in the estimation procedure due to finite sample-size effects. DFA with fourth 
order polynomial detrending is used in order to minimize bias due to local polynomial trends \cite{ya1,jk2}. We also integrate the data prior to DFA estimation
and scale the resulting fluctuation function by the window size \cite{ya1}. The fluctuation function of uncorrelated noise $\alpha=0.5$ and the long-range correlated 
noise $\alpha=0.8~ \rm{and} ~\alpha=0.65$  shall be used as reference in the subsequent discussions.

\subsection{AR (1) with uncorrelated Gaussian innovations}
As noted earlier (Sec. 2) AR (1) process is stationary for $-1<\theta<1$. In the present study, we consider parameters $(\theta=-0.1~ \rm{and}~ 0.1)$
 and $(\theta=-0.93~ \rm{and}~ 0.93)$ . While the former is close to the uncorrelated regime $(\theta=0)$, the latter is close to the non-stationary
 regimes, $\theta=-1~ \rm{and} ~\theta=1$. Log-log plots of the fluctuation function versus time scale for AR (1) and its volatility series with $\theta=0.1 
~\rm{and} ~-0.1$, Fig. 1a, is parallel to exponent 0.5 in the asymptotic regime, with a slight deviation from 0.5 in the short time scales 
for $\theta=-0.1$. A similar analysis was carried for process $\theta=0.93 ~\rm{and}~ -0.93$  Fig. 1b. Unlike $\theta=~\pm~ 0.1$ whose fluctuation function is 
relatively homogenous, characteristic crossovers are observed for $\theta=~\pm~0.93$ and their volatility series, Fig. 1b. For $\theta=0.93$, the log-log plot
of the AR (1) process displayed correlated behavior at short-time scales and uncorrelated behavior at the larger time scales $\alpha \sim 0.5$. However, 
for $\theta=-0.93$ , one observes a characteristic crossover from
 anti-correlated behavior for time scales $(s < 30)$ to uncorrelated behavior
 for $(s > 30)$. The
 volatility series for $\theta=0.93$ was homogenous and exhibited an exponent of $(\alpha\sim 0.5)$ characteristic of monofractal data \cite{ya1}, whereas 
that of $\theta=-0.93$, displayed a marked crossover from correlated to
 uncorrelated behavior. 

{\it From the above discussion, it is clear that the
 process parameter $\theta$ has a significance impact on the scaling behavior
 at shorter time scales. This is reflected by marked distortions and
 cross-overs in the fluctuation plots. It can also be noted that irrespective
 of the choice of $\theta$, the scaling in the asymptotic regime resembles
 that of an uncorrelated noise, $\alpha=0.5$}.

\subsection{AR(1) with long-range correlated Gaussian innovations, CAR (1)}
Long-range correlated Gaussian innovations $\epsilon_n$ with scaling exponent of ($\alpha = 0.8$) \cite{makse} was used to generate CAR (1). The 
fluctuation function of CAR (1) process for $\theta=\pm 0.1$ were parallel to
each other and to $\epsilon_n$  ($\alpha = 0.8$), see Fig. 2a. The corresponding
 volatility series displayed an exponent of 0.5 characteristic of monofractal data. Thus the scaling behavior of CAR (1) is similar to that of AR (1) 
for process parameter $\theta$ close to zero.

For $\theta=-0.93$, at short time scale $s < 35$, CAR (1) is highly anti-correlated however the behavior mimicked that 
of $\epsilon_n$  ($\alpha = 0.8$) for $s > 40$, Fig. 2b. For $\theta = 0.93$, short time scales had an exponent  $\alpha \sim 1.75$ close to the integrated version of $\epsilon_n$ ($\alpha = 0.8$). 
However, in the asymptotic regime the scaling exponent was that of $\epsilon_n$  ($\alpha = 0.8$), Fig. 2b. Fluctuation function obtained for the volatility series ($\theta = 0.93$) 
was parallel to the reference line with slope 0.5 characteristic of monofractal data. However, the fluctuation function of the volatility series ($\theta = -0.93$) displayed 
correlated behavior at short time scales $(s < 200)$ and uncorrelated behavior $(\alpha \sim  0.5)$ in the asymptotic regime. As in the case of AR (1), the process 
parameter $\theta$ has a significant impact on the scaling behavior of CAR (1) process, revealed by marked distortion and crossovers in the log-log plots.

\subsection{Critical note on the choice of CAR (1) to model temperature correlations}
It has been shown \cite{eva1,eva2} that atmospheric temperature data (from a randomly chosen 14 continental regions) display power 
correlations with DFA exponent $\alpha=0.65$ and this is considered as universal persistence behavior of the temperature data. Using this result 
as a benchmark, the quality of the temperature data simulated by the Global Climate Models has been evaluated \cite{rbg}. In a later 
study \cite{janei} it has been shown that temperature data from continental and costal zones display lesser deviations from the previously 
observed universal behavior (exponent) of 0.65 while the islands display greater variability with the average exponent being close
 to 0.8 \cite{roberto}. In a recent study, \cite{jano} CAR(1) process was proposed as a plausible model for explaining correlations in atmospheric 
daily temperature fluctuations collected from fourteen meteorological centers in Hungary. The choice of the CAR (1) model was inspired
 by the fact that temperature records exhibit both short-term and long-term correlations \cite{jano,gas}. We generated the data using the CAR (1)
model proposed by in \cite{jano}
$$x_i=(\alpha_1-c)x_{i-1}+\delta \eta_i.  \eqno(2) $$
The model parameters were chosen as $\alpha_1=0.8, \delta=2.1$, $\eta_i$ is zero-mean, unit variance long-range 
correlated Gaussian innovations with scaling exponent $\alpha=0.65$, $\rho=\alpha-0.5,~c=2\rho^{3/2}$ as in \cite{jano}.

Log-log plot of the fluctuation function of Eqn. 2, Fig. 3, displayed strong correlations with exponent ($\alpha = 0.85$) at short scales $(s < 100)$. However, 
in the asymptotic regime it displayed an exponent $(\alpha\sim  0.65)$ close to that of the long-range correlated Gaussian innovations. Fluctuation function 
of the corresponding volatility series was parallel to $(\alpha = 0.5)$, characteristic of monofractal processes. This has to be contrasted with recent 
studies \cite{gas}, which provided compelling evidence of long-range correlations $(\alpha = 0.6)$ in the volatility series of daily temperature fluctuations. 
Such long-range volatile correlations were observed across randomly chosen and spatially separated meteorological sites across the world. This in 
turn was attributed to universality in daily temperature records \cite{gas}. 

It has been shown that long range correlated data with exponents
$0.5\le \alpha <1.5$ will exhibit uncorrelated behavior in volatility analysis
while the multifractal
signals exhibit long range correlations in volatility
analysis \cite{ya1,ay2}.  As the CAR (1) model,
Eqn. 2, is driven by a linear process (see above), one would expect the model
to display uncorrelated behavior in the volatility analysis and is indeed the
case obtained for this model. Since, the CAR (1) model Eqn.2 \cite{jano} captures the linear features of the
 temperature data (as the volatility exponent is 0.5 for this model), it will
 be unable to account for the nonlinear features in the data irrespective of
 the choice of the process parameter, as demonstrated  
by Figs. 2 and 3. Equation (2) \cite{jano} does show distinct scaling behavior with a marked crossover, however, it is not appropriate for modeling the
 dynamics of daily temperature records. In \cite{jano}, the authors subsequently proposed a nonlinear counterpart of CAR(1) model, termed 
NLCAR(1) in order to capture the probability distribution of the daily
temperature records. The DFA results obtained in \cite{jano} for NLCAR(1) and CAR(1) were
 indistinguishable (for details we refer to \cite{jano}). However, this
 does not necessarily imply NLCAR(1) and CAR(1) have similar volatile
 correlations. Recent studies 
 \cite{ay3} have 
 used nonlinear stochastic models successfully to model glacial dynamics also
 reflected in long-range volatile correlations. Encouraged by previous findings
 \cite{ya1,gas,ay3} and results presented in \cite{jano}, we believe that
 NLCAR(1) may be a more appropriate model to capture dynamics of daily
 temperature records.

\newpage

\begin{figure}[H] 
  \begin{center} 
    \includegraphics[width=5in,height=3.5in,angle=0]{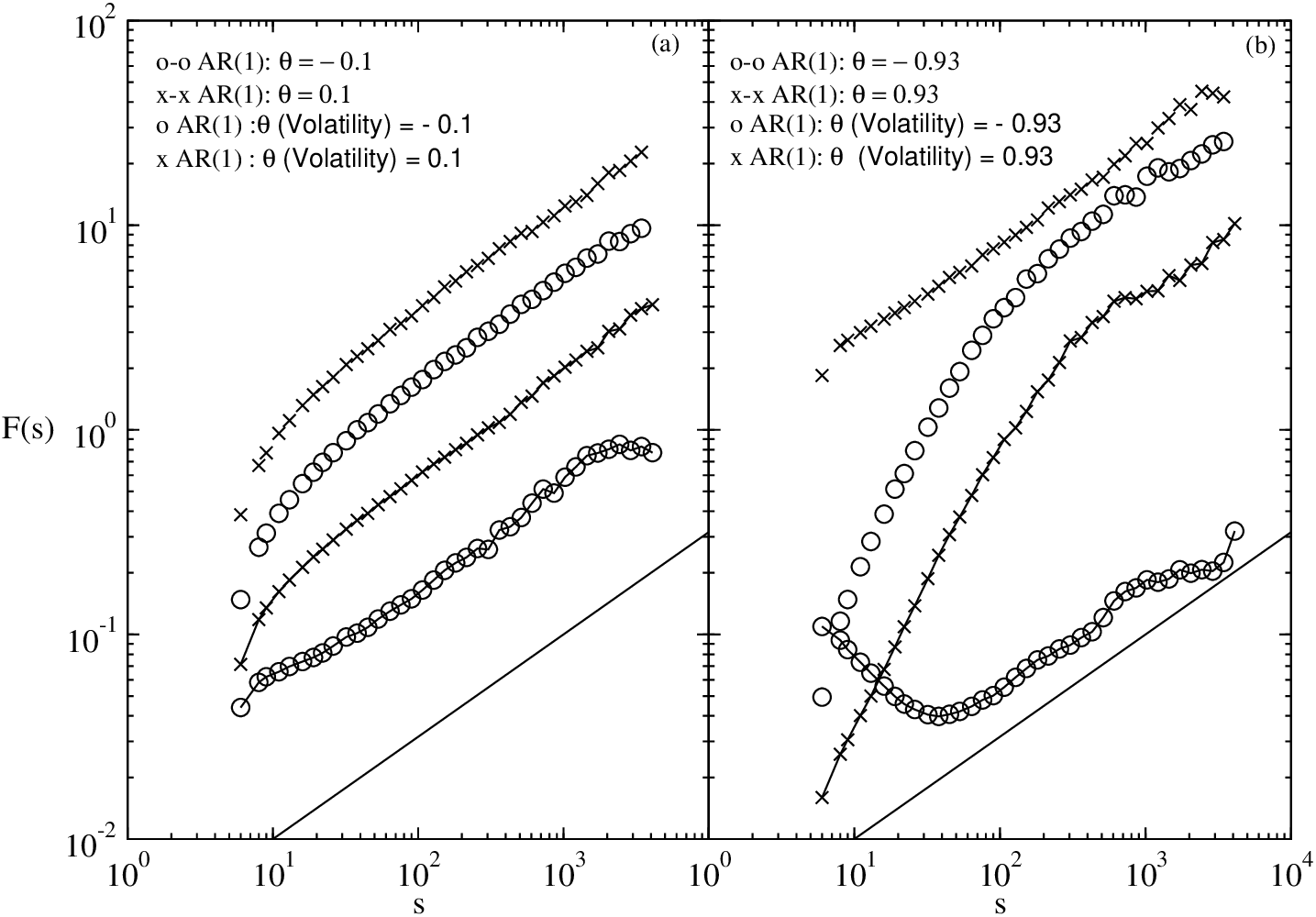} 
    \caption{Scaling analysis of AR (1) process. (a) Fluctuation function for $\theta=-0.1$ and $\theta=0.1$ is shown by (solid curve with open circles) 
and (solid curve with \upshape{x}) respectively. Corresponding volatility analysis for $\theta=-0.1$ and $\theta=-0.1$ are shown
with (open circles and \upshape{x}) respectively. (b) Fluctuation function for $\theta=-0.93$ and $\theta=0.93$ is shown
 by (solid curve with open circles) and (solid curve with \upshape{x}) respectively. Corresponding volatility analysis 
for $\theta=-0.93$ and $\theta=0.93$ are shown with (open circles and \upshape{x}) respectively. Fluctuation function of
the uncorrelated noise with slope 0.5 is shown at the bottom of Fig 1 (a) and (b) as reference. In the asymptotic regime all 
the curves are parallel to the reference line indicating uncorrelated nature
of the process.} 
  \label{fig1} 
  \end{center} 
\end{figure}

\begin{figure}[H]
  \begin{center} 
    \includegraphics[width=5in,height=3.5in,angle=0]{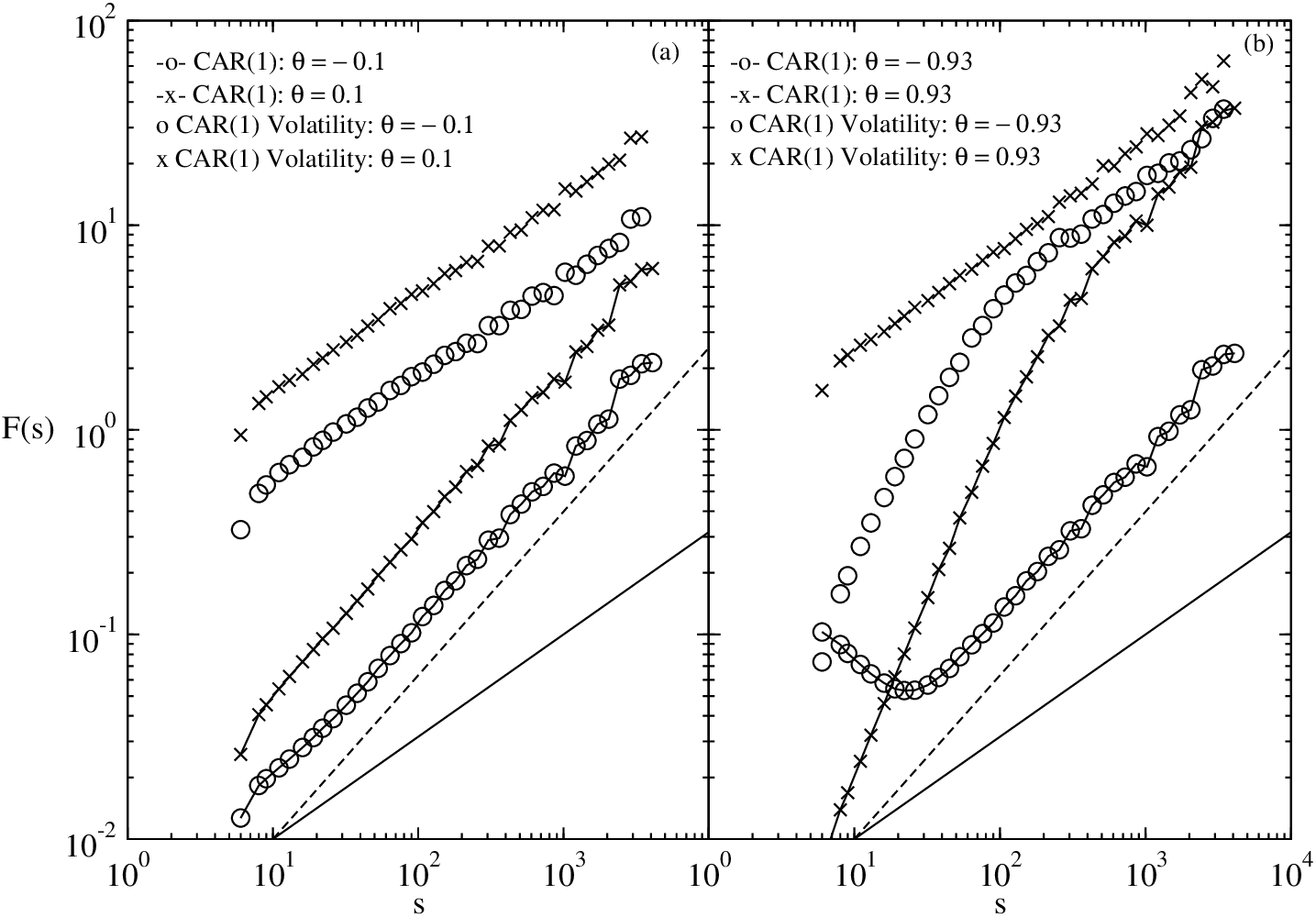} 
    \caption{Scaling analysis of CAR(1) process (i.e. AR(1) process driven by long range 
correlated noise with DFA exponent 0.8) for process parameters ($\theta=\pm0.1$ and $\theta=\pm0.93$) and their 
volatility series is shown in (a) and (b) respectively. Explanations are same as in Fig. 1. Fluctuation function of the uncorrelated 
noise (slope 0.5, solid line) and long-range correlated noise (slope = 0.8, dotted line) is shown at the bottom in (a) and (b) as reference. 
In the asymptotic regime DFA curves for the original data are parallel to the reference line with slope 0.8 and the DFA
 curves of the volatile data are parallel to the reference to line with slope 0.5. }
  \label{fig2} 
  \end{center} 
\end{figure}

\begin{figure}[H]
  \begin{center} 
    \includegraphics[width=5in,height=5in,angle=0]{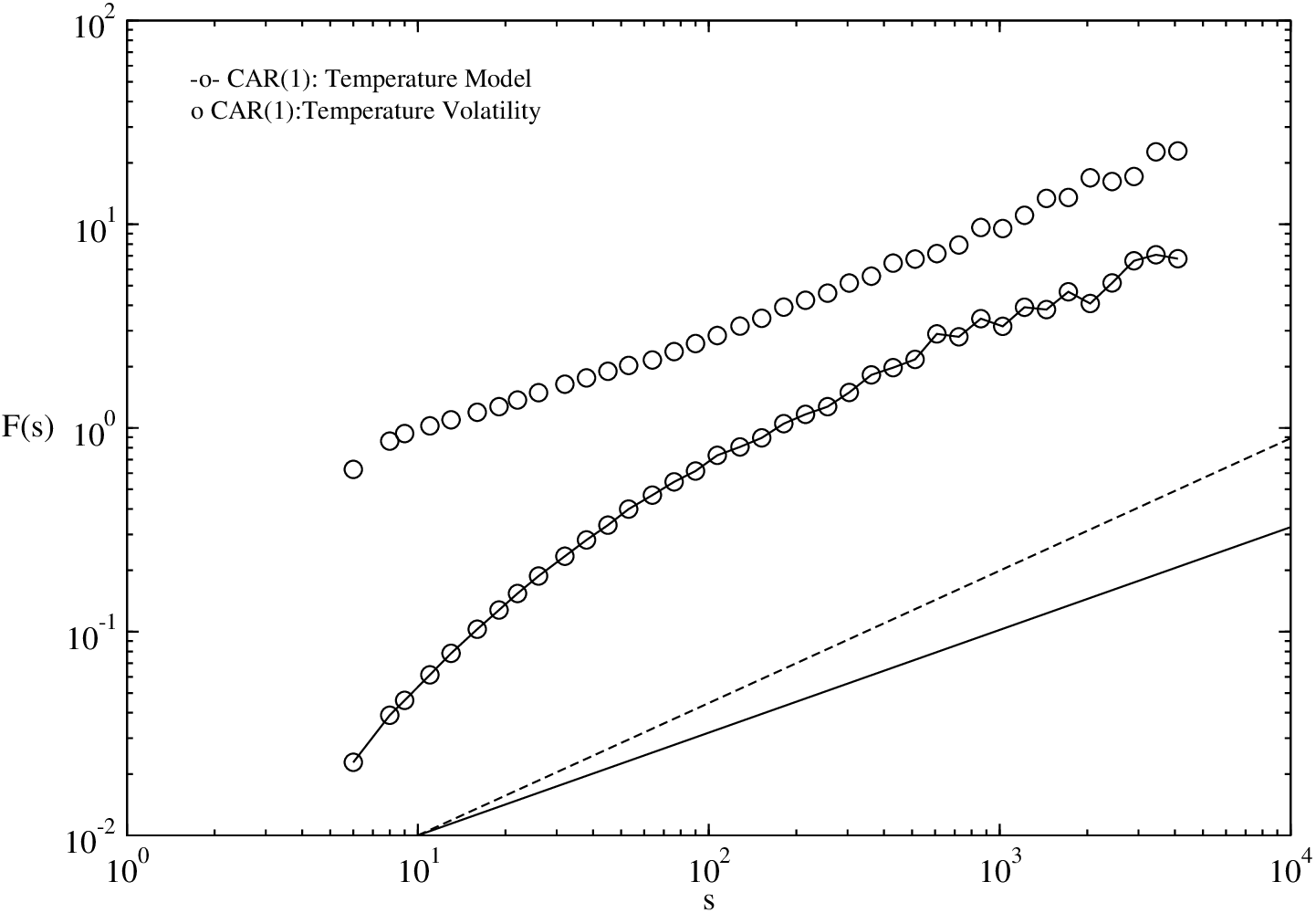} 
    \caption{Scaling analysis of CAR(1) temperature model (Sec. 3, Eqn. 2). The fluctuation function for the temperature model is shown
 in solid curve with open circles. The fluctuation function of the corresponding volatility series is shown with open circle. Fluctuation 
function of the uncorrelated noise (slope 0.5, solid line) and long-range correlated noise (slope = 0.65, dotted line) is shown at the 
bottom in (a) and (b) as reference. In the asymptotic regime, DFA curve of the original data is parallel to the reference line with 
slope 0.65 while at short time scales $s<200$ the data are correlated with an exponent $\alpha=0.85$. The DFA curves of the volatile
data are parallel to the reference line with slope 0.5 indicating uncorrelated nature of the volatile correlations.} 
  \label{fig3} 
  \end{center} 
\end{figure} 
\end{document}